\documentstyle[aps,psfig,prl]{revtex}
\newcommand{\avg}[1]{\langle{#1}\rangle}
\newcommand{\req}[1]{(\ref{#1})}
\newcommand{\beq}{\begin{equation}}
\newcommand{\eeq}{\end{equation}}
\newcommand{\beqar}{\begin{eqnarray}}
\newcommand{\eeqar}{\end{eqnarray}}
\begin{document}
\twocolumn[\hsize\textwidth\columnwidth\hsize\csname
@twocolumnfalse\endcsname
\title{Critical exponents of the anisotropic Bak-Sneppen model}
\author{Sergei Maslov$^{(1)}$, Paolo De Los Rios$^{(2)}$, 
Matteo  Marsili$^{(2,3)}$, and Yi-Cheng Zhang$^{(2)}$}
\address{$^1$Department of Physics, Brookhaven National
Laboratory, Upton, New York 11973, USA}
\address{$^2$Institut de Physique Th\'eorique,
Universit\'e de Fribourg P\'erolles, Fribourg CH-1700, 
Switzerland}
\address{$^3$International School for Advanced Studies (SISSA)
and INFM unit, Trieste I-34014, Italy}

\date{\today}
\maketitle
\widetext
\begin{abstract}
We analyze the behavior of spatially anisotropic Bak-Sneppen model. 
We demonstrate that a nontrivial relation between critical exponents 
$\tau$ and $\mu=d/D$, recently derived for the isotropic Bak-Sneppen model, 
holds for its anisotropic version as well. For 
one-dimensional anisotropic Bak-Sneppen model we 
derive a novel exact equation for the distribution of 
avalanche spatial sizes, and extract the value 
$\gamma=2$ for one of the critical exponents of the model. 
Other critical exponents are then determined from 
previously known exponent relations.
Our results are in excellent agreement with 
Monte Carlo simulations of the model as well
as with direct numerical integration 
of the new equation.
\end{abstract}
\pacs{05.40+j, 64.60Ak, 64.60Fr, 87.10+e}
]
\narrowtext

Ever since its introduction five years ago, the Bak-Sneppen (BS) model
\cite{bs} was a subject of considerable theoretical interest. 
Its relevance relies on the fact that 
it provides a very simple mechanism of Self-Organized Criticality 
\cite{BTW}.
In fact the Bak-Sneppen model is the simplest representative of a
broad class of extremal models, which all 
naturally evolve towards a scale-free stationary state \cite{PMB}.
An extremely rich dynamic critical behavior arising out of 
truly minimalistic dynamical
rules has inspired numerous analytical and numerical  
investigations of the Bak-Sneppen model 
\cite{PMB,mf,BSRG,backw,multitrait,master,MDM,hdbs}. 

In this work we introduce and study an {\it anisotropic} version of the 
original Bak-Sneppen model. In one dimension the dynamics 
is as follows: the configuration of the system is fully defined 
by the value of variable 
$f_i$ for each lattice site $i$. At every
time step the smallest variable in the system and that 
at its {\it right} nearest neighbor are replaced with new random 
numbers independently drawn from the distribution 
${\cal P}(f)=e^{-f}$ \cite{notap}. Contrary to the 
isotropic BS model, where both nearest neighbor variables are 
updated, only the variable in the preferred direction 
is updated here. Other mechanisms of
introducing spatial anisotropy to the rules of the original BS model 
were recently studied in Refs. \cite{aniBS1,aniBS2}. 
As we shall see, the universality of the 
critical behavior manifests itself in the fact that any realization of
the anisotropy in the original BS model gives rise to the same 
set of critical exponents \cite{aniBS1}. 
The generalization of our version of anisotropic BS model 
to higher dimensions is 
straightforward: only $d$ neighbors of the global minimal site located 
in positive directions of corresponding coordinates are updated along 
with it.

This work is devoted to analytical and numerical study of 
exponents of the anisotropic Bak-Sneppen model. 
The main observations used in the analytical part of this 
study are: 
{\em i)} the general scaling theory of Ref.\cite{PMB}
developed for an arbitrary extremal model reduces the number of 
independent critical exponents to just two; 
{\em ii)} the relation $\tau(\mu)$ between two remaining 
exponents $\tau$ and $\mu=d/D$, recently derived in \cite{master,MDM}
for the isotropic Bak-Sneppen model in an arbitrary dimension, 
holds for its anisotropic version as well. 
%This reduces the number of independent critical exponents to
%just one. 
Finally, {\em iii)} in the 
one-dimensional anisotropic BS model the exponent $\gamma=2$, 
describing the divergence of the average avalanche size, can be
derived from the {\it exact} master equation for the 
probability distribution of avalanche spatial sizes. 
This equation, which we
derive in this paper, compliments that for the probability 
distribution of avalanche temporal durations \cite{master}. 
From the exact result $\gamma=2$ it follows 
that the values of exponents $\tau$ and $\mu$ in the 
one-dimensional anisotropic BS model are related via $\tau=2\mu$.
Using the approximate form of the function $\tau(\mu)$, 
derived in \cite{MDM}, we can give an analytical estimate
of the exponents $\tau$ and $\mu$ in 1D anisotropic BS model. 
Indeed, they must lie at the 
intersection point of the $\tau=2\mu$ line and 
the  $\tau(\mu)$ curve, 
valid for an arbitrary BS model \cite{MDM}. 
These analytical estimates are in excellent agreement 
with the direct numerical integration of the exact 
master equation for PDF of the avalanche spatial sizes,  
giving $\mu=0.588(1)$ and $\tau=1.176(2)$,  
as well as Monte Carlo simulations performed by us and 
by he authors of Refs. \cite{aniBS1,aniBS2}.

The scale-free stationary state of an arbitrary extremal model can 
be characterized by a number of critical 
exponents. Several scaling relations, reviewed and discussed
in Ref. \cite{PMB}, reduce this variety to only two independent 
exponents such as  $\tau$ for the power law  
in the distribution $P(s) \sim s^{-\tau}$
of avalanche temporal durations $s$, and the dynamic exponent
$\mu$. The latter exponent relates avalanche temporal 
duration $s$ to its spatial volume $V(s)$ as $V(s)\sim s^\mu$.
Here the spatial volume is defined as the number of distinct 
sites updated at least once during this avalanche. In the notation
of Ref. \cite{PMB} $\mu=d/D$ (or $\mu=d_{\rm cov}/D$ if the set 
of updated sites is not compact, but instead forms a fractal of 
dimension $d_{\rm cov}$).

The Bak-Sneppen model in an arbitrary dimension and with an 
arbitrary anisotropy has an additional simplification 
\cite{PMB}: since 
variables are simply replaced with new random numbers and have no
memory about their previous values, the dynamics within a 
single avalanche is totally independent 
from what happened before it started. 
This observation \cite{PMB} enables one to simulate 
$f$-avalanches (for definitions see \cite{PMB}) 
for an arbitrary value 
of $f$, which can be above as well as below the critical point, 
without specifying variables at passive sites (those with $f_i>f$) 
prior to avalanche. Another important consequence of the absence 
of inter-avalanche memory is that in any variant 
of BS model there are no correlations between sizes of subsequent 
avalanches. This statement is {\it not} a result of any kind of 
mean-field approximation. Rather it 
is a clear logical consequence of the dynamical rules of 
the Bak-Sneppen model.  

In \cite{master} one of us derived a 
master equation for the distribution $P(s,f)$ of avalanche durations
$s$, valid for a general BS model. 
Let us recall briefly the sequence of logical steps leading to this 
equation. 
The starting point is the analysis of the signal (minimal number 
as a function of time) $f_{\min}(t)$  of the model using an auxiliary 
parameter $f$. The intersection of this signal with the horizontal
line drawn at $f$ identifies the sequence of $f$-avalanches, 
following one another. In other words, if 
$f_{\min}(t)>f$, $f_{\min}(t+k)<f$ for $1\le k< s$, and 
$f_{\min}(t+s)>f$, we say that an $f$-avalanche of size 
(temporal duration) $s$ has 
occurred \cite{PMB}. The sequence of $f$-avalanches is characterized by the 
distribution $P(s,f)$ of their sizes $s$. One can investigate how 
this distribution changes under an infinitesimal increase of $f$ from $f$ 
to $f+df$. The change occurs simply because when the horizontal line at $f$
is lifted, some intersections with the signal $f_{\min}(t)$ disappear. 
This means 
that two consecutive $f$-avalanches of (temporal) sizes $s_1$ and $s_2$  
merge into a single $f+df$-avalanche of size $s_1+s_2$. The occurrence 
of this event implies that at least one of the $V_1 \sim s_1^\mu$ sites,  
updated during the course of the first avalanche, 
has $f<f_i<f+df$. Taking into account that as we argued above 
subsequent avalanches in Bak-Sneppen
model are uncorrelated, one can write the balance 
of loss and gain of $f$-avalanches of size $s$ as $f$ is 
increased. The resulting master equation for the distribution $P(s,f)$ 
is given by 
\beq
\partial_f P(s,f) = -s^{\mu} P(s,f)+ 
\sum_{s_1=1}^{s-1} s_1^{\mu} P(s_1,f) P(s-s_1,f).
\label{s_hier}
\eeq
Strictly speaking in order for the above equation 
to be {\it exact} one needs to replace $s^{\mu}$ with 
the average number of updated sites $V(s,f)$, where the average is taken 
over all $f$-avalanches of size $s$. As was observed numerically this 
quantity has an insignificant $f$-dependence and its large $s$ 
asymptotics is well described by the power law $V(s,f) \simeq A
s^{\mu}$. It was suggested in \cite{master} and later 
convincingly confirmed numerically in \cite{MDM} that 
the critical exponent $\mu$ uniquely determines the scaling 
properties of $P(s,f)$. This justifies our substitution of $V(s,f)$ 
by its asymptotical form $s^{\mu}$ in Eq. \req{s_hier} 
(the constant $A$ in front of 
$s^{\mu}$ was absorbed by redefinition of $f$).

The numerical integration of equation \req{s_hier} with initial
conditions $P(s,f=0)=\delta_{s,1}$ shows that as 
$f$ approaches some critical value $f_c(\mu)$, 
$P(s,f)$ develops a power law form with a 
diverging cutoff: $P(s,f)=s^{-\tau}F(s^{\sigma}(f_c-f))$.
Above $f_c$ there is a finite probability $p_{\infty}(f)$ 
to start an avalanche that never ends. The possibility 
of such event shows up in Eq. \req{s_hier} through 
the ``normalization catastrophe'', when 
$\sum_{s=1}^{\infty} P(s,f)$ for $f>f_c$ starts 
to fall below unity. This deviation is attributed 
to the appearance of the ``infinite avalanche'' 
with probability $p_{\infty}(f) \sim (f-f_c)^{\beta}$. 
This way the overall normalization 
$\sum_{s=1}^{\infty}P(s,f)+p_{\infty}(f)=1$ is satisfied at all $f$.   
The properties of  Eq. \req{s_hier} depend on the 
critical exponent $\mu$.
Given the value of this dynamic critical exponent, 
the remaining exponents $\tau$, $\sigma=\mu+1-\tau$, 
$\gamma=(2-\tau)/\sigma$, and $\beta= (\tau-1)/\sigma$, 
as well as the scaling function $F(x)$ of the
avalanche distribution near the critical point 
follow from Eq. \req{s_hier}. In \cite{MDM} the 
function $\tau(\mu)$ and the scaling form $F(x)$ 
were determined by numerical integration of Eq. \req{s_hier} 
and the expansion around the mean-field point $\mu=1$, $\tau=3/2$.
It was shown \cite{MDM} that to a second order in $1-\mu$ the function
$\tau(\mu)$ is given by 
\beqar
\tau(\mu) &=& 1.5-(1-\mu)+c(1-\mu)^2+O((1-\mu)^3);
\nonumber \\
c&=&{4 \over 3} (\gamma_e+\ln 2-1) \simeq 0.3605 .
\label{tau}
\eeqar 
Here $\gamma_e \simeq 0.5772$ is the Euler's constant.  
%The quality of this approximate formula is quite satisfactory
%everywhere except for very small $\mu$.  

Our new results for the one-dimensional anisotropic BS model are 
based on the exact equation for the probability 
distribution function $Q(r,f)$ of {\it spatial} sizes $r$ of 
$f$-avalanches. The derivation of this equation is very similar to the
derivation of \req{s_hier}, briefly outlined
above. The main difference lies in the fact that when an avalanche of
spatial size $r_1$ merges with that of size $r_2$ they can overlap to 
form any spatial size between $\max (r_1,r_2)$ and $r_1+r_2-1$.
Contrary to this, in the temporal domain the merging of avalanches 
of temporal durations $s_1$ and $s_2$ always produces an avalanche 
of temporal duration $s_1+s_2$. Let us analyze how $Q(r,f)$ changes 
when $f$ is increased by an infinitesimal amount $df$. 
Some avalanches of size $r$ 
will merge with the next avalanche. This event 
can only occur if one of the $r$ sites, updated in the first 
avalanche, happens to host 
the smallest number right after the avalanche is finished. 
Each of these $r$ sites has a variable $f_i$, which was randomly drawn 
from ${\cal P}(f)=e^{-f}$ during the 
course of the avalanche. At the end of this avalanche  
by the very definition of an $f$-avalanche all these $r$ 
sites have $f_i>f$. We can therefore regard 
the $f_i$'s on these sites as randomly drawn from an exponential 
distribution normalized between $f$ and $\infty$. The probability that a 
particular $f_i$ is in the interval $[f,f+df]$, to linear order in $df$, 
is just $df$. The probability that at least one of the $r$ sites has an 
$f_i$ in the interval $[f,f+df]$ is $r df+O(df^2)$. This implies 
that the number of $f$-avalanches which will merge with 
the next one when $f$ is raised 
to $f+df$ is 
%\[
$dQ(r)|_{\rm loss}=-rdf Q(r,f)+O(df^2)$.
%\]
Let us now consider a merging event between two
$f$-avalanches of size $r_1$ and $r_2$ resulting in an
$f+df$-avalanche of size $r$.  
There are two scenarios of how this can happen:
{\em(i)} the rightmost point of the second avalanche 
is at distance $r$ from the leftmost (starting) 
point of the first avalanche. The constraint on possible 
values of $r_1$ and $r_2$ imposed by this scenario
is $\max (r_1,r_2) \leq r \le r_1+r_2-1$;
{\em(ii)} $r_1=r$, and the second avalanche is fully contained 
within the first one. In the former case, the values of 
$r$, $r_1$, and $r_2$ uniquely specify the initial site of the
second avalanche. Therefore, the probability of this event to occur 
is just the probability $df+O(df^2)$ that this site has $f_i\in[f,f+df]$.
However, in the latter case the starting point 
of the second avalanche can 
be any of the first $r-r_2$ sites of the first avalanche. 
This event occurs with a probability $(r-r_2)df+O(df^2)$. 
Putting all these terms together we find: 
\beqar
\partial_f Q(r,f)&=& -r Q(r,f)+ 
\sum_{r_1=1}^r Q(r_1,f)\sum_{r_2=r-r_1+1}^r Q(r_2,f) \nonumber \\
&+&Q(r,f) \sum_{r_2=1}^{r} (r-r_2) Q(r_2,f).
\label{m_ani}
\eeqar

This is an {\it exact} equation for the distribution 
of spatial sizes of $f$-avalanches. Unlike our previous results,  
its validity does not require any scaling assumptions, 
such as $V(s,f) \sim s^{\mu}$ used in the derivation of Eq. 
\req{s_hier}.  The equation \req{m_ani} has to be solved
with the initial condition $Q(r,f=0)=\delta_{r,2}$. 
Indeed, in the one-dimensional anisotropic BS model at any time step 
(or $f=0$-avalanche for that matter) $r=2$ sites are updated.

A similar but more complicated equation can be written
for the {\it isotropic} one-dimensional Bak-Sneppen model. 
The basic object of this equation
is the probability distribution $Q(r_1,r_2,f)$ 
of $f$-avalanches replacing precisely $r_1$ sites to the left of
the starting point, and $r_2$ sites to the right of it. The initial
condition is given by $Q(r_1,r_2,0)=\delta_{r_1,1} \delta_{r_2,1}$.
Although we were unable to derive any analytical exponent relations 
from this equation, it should be stressed that it 
contains all properties of the one-dimensional 
isotropic BS model, and in principle its numerical integration
constitutes a viable alternative to Monte Carlo simulations of the 
model.

Yet another variant of Eq. \req{m_ani} can be written for 
the anisotropic Bak-Sneppen model in dimensions higher than one. 
Let $r$ to denote the spatial extent of the avalanche, 
measured along the diagonal $(1,1, \ldots, 1)$ of the 
$d$-dimensional space. In projection to this axis 
the starting point of an avalanche is
always the leftmost point in the avalanche. In order to describe the
shape of the region covered by avalanche one needs to introduce the 
new exponent $\zeta$, defined by the average number of updated sites  
$n_{\rm proj}(r) \sim r^{\zeta}$ projected onto the same point 
at the distance $r$ along the diagonal from the starting point. 
Then the total number of points covered by an avalanche of 
size $r$, $n_{\rm cov}(r) \sim 
n_{\rm proj}(r) \times r \sim r^{1+\zeta}$. In 1D anisotropic BS model 
$n_{\rm proj}(r)=1$, and, therefore, $\zeta=0$. By retracing 
arguments that led to the derivation of Eq. \req{m_ani} 
it is easy to see that its higher-dimensional version
can be written as
\beqar
\partial_f Q(r,f)&=& -r^{1+\zeta} Q(r,f) \nonumber \\ 
&+&\sum_{r_1=1}^{r} r_1^{\zeta} Q(r_1,f)\sum_{r_2=r-r_1+1}^r Q(r_2,f) 
\nonumber \\ 
&+&Q(r,f) \sum_{r_s=1}^{r} r_s^{\zeta} \sum_{r_2=1}^{r-r_s-1} Q(r_2,f).
\label{m_ani_high_d}
\eeqar
The drawback of this equation is that, similar to Eq. \req{s_hier},  
it requires the input of an additional parameter (critical exponent) 
$\zeta$.

Note that the Equation \req{m_ani} for $Q(r,f)$ involves only
$Q(r',f)$ for $r'\le r$. Therefore, in principle this distribution
can be computed numerically for $r\le R$ to the desired 
accuracy. Forward numerical integration of Eq. \req{m_ani} 
shows that as $f\to f_c \approx 1.2865$, $Q(r,f)$ develops a power
law behavior with an exponent $\tau_r=1.299(3)$ (see Fig. 1). 
It is easy to see that in one dimension 
the exponent $\tau_r$ of the distribution 
of avalanche spatial sizes is related to the 
more familiar exponent $\tau$ of the distribution of 
their temporal durations through $\tau_r=(\tau-1)/\mu+1$. 
Indeed, since asymptotically $r=As^{\mu}$, $s_0^{1-\tau} \sim 
P(s>s_0)=P(r>A s_0^{\mu}) \sim (A s_0^{\mu})^{1-\tau_r}$. 
Comparing powers of $s_0$ in this expression one gets the 
above exponent relation. 
To find values of critical exponents 
hidden in Eq. \req{m_ani} we study the behavior 
of moments $\avg{r^n}$ of the distribution $Q(r,f)$ as a 
function of $f$. These satisfy the equation:
\beqar
\partial_f &&\avg{r^n} = 
-\avg{r^{n+1}}+\sum_{r_{+}=1}^\infty\sum_{r_{-}=1}^{r_{+}} 
\nonumber \\
&&\left[2\sum_{\rho=r_{+}}^{r_{+}+r_{-}-1} \rho^n+ r_{+}^n (r_{+}-r_{-})\right]Q(r_{+},f)Q(r_{-},f).
\label{avg}
\eeqar
where the sum over $r_1$ and $r_2$ has been 
transformed into a sum over $r_{+}=\max(r_1,r_2)$
and  $r_{-}=\min(r_1,r_2)$. For $n=1$ the Eq. \req{avg} reads 
$%$
\partial_f 
\avg{r}=-\avg{r^{2}}+\sum_{r_{+}=1}^\infty\sum_{r_{-}=1}^{r_{+}} 
(r_{+}^2+r_{-}^2+r_{-}r_{+}-r_{-}) Q(r_{+},f) Q(r_{-},f)=-\avg{r^{2}}+
\sum_{r_1=1}^\infty \sum_{r_2=1}^\infty
[r_1^2+r_2^2+r_1 r_2 -\min (r_1, r_2)]/2\ Q(r_1,f) Q(r_2,f)
$. %$ 
For $f<f_c$, when there are no infinite avalanches and 
the avalanche distribution 
$Q(r,f)$ is normalized to unity,  one gets
\beq
\partial_f \avg{r}={\avg{r}^{2}\over 2}-{\avg{\min(r_1, r_2)} \over 2}.
\label{avg1}
\eeq
Close to the critical point we can neglect 
the term $\avg{\min(r_1, r_2)}$, since it diverges slower 
than $\avg{r}^2$ and we are left with
$\partial_f \avg{r}\simeq \avg{r}^{2}/2$.
This has an obvious solution
\beq
\avg{r}=\frac{2}{f_c-f}+O\left({1 \over (f_c-f)^{2}}\right).
\label{avg2}
\eeq 
It has been known for some time that in the Bak-Sneppen model as well as
in several other extremal models 
the {\it amplitude} of the divergence of 
$\avg{r}$ as $f\to f_c^-$ is given 
by the critical exponent $\gamma$. 
This exponent describes the critical behavior of the average avalanche 
duration as $\avg{s} \sim |f_c-f|^{-\gamma}$. 
For the original derivation of this fact \cite{remark} 
see Eq. (17) in \cite{PMB}. Actually, as it was shown 
in \cite{master}, this fact can be also derived 
from Eq. \req{s_hier}. Indeed, from this equation 
it is easy to see that  
in the absence of infinite avalanche 
the first moment of $P(s,f)$ obeys 
$\partial _f \avg{s} = \avg{s^{\mu}} \avg{s}$. Therefore, 
in the critical region one has 
$\avg{r}=\avg{s^{\mu}}=\partial_f(\ln \avg{s})=
\gamma/(f_c-f)+O((f_c-f)^{-2})$!
From Eq. \req{avg2} we conclude that in the 
anisotropic one-dimensional Bak-Sneppen model  
\beq
\gamma=2.
\label{gamma}
\eeq

Using the scaling relation \cite{PMB} 
$\gamma=(2-\tau)/\sigma=(2-\tau)/(1+\mu-\tau)$, we readily find
that Eq. \req{gamma} implies $\tau=2\mu$.
This, combined with the $\tau(\mu)$ relation 
found in \cite{MDM},  gives $\mu=0.58(1)$ and
$\tau=1.16(2)$ as coordinates of the intersection 
point (see Fig. 2).  The uncertainty in these numbers comes 
from our approximate knowledge of the systematic errors in the 
$\tau(\mu)$ curve, determined by numerical integration of 
Eq. \req{s_hier}. Presently, this integration was performed 
and approximate power law exponent $\tau$ was measured 
for $P(s,f)$ with $s \leq 2^{14} \simeq 1.6 \times 10^4$. 
To improve the precision we can use a very accurate 
value for $\tau_r=1+(\tau-1)/\mu=1.299(3)$ measured 
by direct numerical integration of Eq. \req{m_ani}.
Our resources allowed us to integrate this 
equation forward for $r \leq 2^{14}$, which 
is equivalent to measuring $P(s,f)$ up to 
$s=r^{1/\mu} \simeq 1.5 \times 10^7$, i.e. over much 
wider range than from numerical integration of Eq. \req{s_hier}.
Using exponent relations $\tau=2\mu$ and $\tau=1+\mu(\tau_r-1)$ 
from  $\tau_r=1.299(3)$ one gets
\beqar
\tau=1.176(2) \qquad ;\\
\mu=0.588(1) \qquad .
\eeqar
These are our best estimates of two basic exponents of the 
one-dimensional anisotropic BS model. 
The Monte Carlo simulations of the 
anisotropic BS model in $d=1$ are in perfect agreement 
with these values for $\tau$ and $\mu$. 
Indeed, Head and Rodgers \cite{aniBS2}
found $\mu=0.59(3)$, while in \cite{aniBS1} it was measured 
to be $\mu=0.60(1)$. 
In \cite{aniBS2} they also measured the exponent 
$\pi=2.42(5)$ of the distribution 
of spatial jumps of the minimal site. This should be compared to our
prediction for this exponent based on the scaling relation 
$\pi=1+(2-\tau)/\mu=2.401(6)$. 
We also have performed Monte Carlo simulations of the 
anisotropic BS model in one and two dimensions. In 1D we found 
$\mu_{1D}=0.58(1)$ and $\tau_{1D}=1.17(1)$ in agreement with 
\cite{aniBS1,aniBS2}, our analytical results, and 
the direct simulation of Eq.\req{m_ani}. 
In two dimensions our Monte Carlo simulations 
give $\mu_{2D}=0.83(1)$ and $\tau_{2D}=1.35(1)$. 
As shown in Fig. 2 these exponents lie on the $\tau(\mu)$ 
curve, valid for a general BS model. However, 
in the 2D anisotropic BS model we don't know 
the exact value of $\gamma$ to fix the 
position of the exponents on this curve.  

We also tested our theoretical prediction 
$\avg{r} \simeq 2/(f_c-f)$ (which led us to $\gamma=2$) 
against the direct numerical integration of Eq. \req{m_ani}.
We measured the first moment 
$\avg{r}$ of the distribution $Q(r,f)$, obtained 
as described above by direct numerical integration of
Eq. \req{m_ani}. As expected the power law divergence of $\avg{r}$ 
both above and below $f_c$ has the exponent $-1$. It can be clearly 
seen when numerical derivative $d(\avg{r}^{-1})/df$ is plotted 
as a function of $f$ (see Fig. 3). 
This derivative approaches different {\it finite} limits as 
$f \to f_c \pm 0$. The divergence of 
$\avg{r}$ at
$f_c$ is clearly cut off by the finite size $R$ of
avalanches considered. This is illustrated in Fig. 3 
by showing two curves for two different cutoff sizes $R=2^{12}$ and 
$2^{14}$. From the asymptotical value of $d(\avg{r}^{-1})/df$ 
as $f \to f_c-0$ we measure 
$1/\gamma =0.498(3)$, which 
yields $\gamma=2.01(2)$, in complete agreement 
with Eq. \req{gamma}.

We were able to estimate yet another 
critical exponent from this plot. As it was shown in \cite{master}, 
the divergence of $\avg{r}$ in the overcritical regime 
{\it above} $f_c$ is given by $\avg{r}=\beta/(f-f_c)$, 
where $\beta=(\tau-1)/(1+\mu-\tau)$ is the ``order parameter''
exponent describing the scaling of the probability to start an
infinite avalanche $p_{\infty} (f) \sim (f-f_c)^{\beta}$. 
For $p_{\infty}=1-\sum_{s=1}^{\infty} P(s,f)$ 
the Eq. \req{s_hier} readily gives
$\partial _f  p_{\infty} = \avg{s^{\mu}} p_{\infty}$. 
Therefore, above the critical point one has 
$\avg{r}=\avg{s^{\mu}}=\partial _f (\ln p_{\infty})=
\beta/(f-f_c)+O((f-f_c)^{-2})$! 
The quadratic fit to $d(\avg{r}^{-1})/df$ above $f_c$ gives
$1/\beta=2.34(2)$, or $\beta=0.427(4)$. This is
in excellent agreement with our theoretical prediction 
$\beta=(\tau-1)/(1+\mu-\tau)=0.427(6)$ based on 
the best estimate of $\tau_r=1.299(3)$ and the 
exponent relation $\beta=(\tau_r-1)/(2-\tau_r)$.

In summary, we have analyzed the behavior of the
anisotropic Bak-Sneppen model. We demonstrated that a nontrivial 
relation between critical exponents $\tau$ and $\mu$, recently derived
for the isotropic Bak-Sneppen model \cite{master,MDM}, 
holds for its anisotropic version as well.
The exponents measured by Monte Carlo simulations of 
the anisotropic Bak-Sneppen model in one and two dimensions 
are in agreement with this relation.  For 
one-dimensional anisotropic Bak-Sneppen model we 
derive a novel exact equation \req{m_ani} 
for the distribution $Q(r,f)$ of avalanche spatial sizes. 
We also propose analogous equations for one-dimensional isotropic
BS model and anisotropic BS model in $d>1$. By studying the behavior 
of the first moment of the distribution $Q(r,f)$ 
we managed to extract the exact value $\gamma=2$ for one of 
the critical exponents of the one-dimensional anisotropic BS model. 
The values of critical exponents $\tau$ and 
$\mu=d/D$ were found as coordinates of the 
intersection point between $\tau(\mu)$ and 
$\gamma (\tau, \mu)=2$ curves. They are in excellent agreement with
both Monte Carlo simulations of the model as well as results 
of numerical integration of the master equation for $Q(r,f)$. 
We summarize our best estimates for the exponents in one and two 
dimensions in Table 1.

One of us (S.M.) would like to thank N.D.Mermin for 
asking the question that triggered this study. 
The work at Brookhaven National Laboratory was 
supported by the U.S. Department of Energy Division
of Material Science, under contract DE-AC02-98CH10886.

\newpage

\begin{table}
\begin{tabular}{ccc}
Exponent & 1D anisotropic BS & 2D anisotropic BS \\ \hline
$\tau$ & 1.176(2) & 1.35(1) \\
$\mu$  & 0.588(1) & 0.83(1) \\
$\sigma$ & 0.412(1) & 0.48(2) \\
$\gamma$ & 2 & 1.35(4) \\
$\beta$  & 0.427(6) & 0.73(4) \\
$D$ & 1.701(3) & 2.41(3) \\
$\pi$ & 2.401(6) & 2.57(3)
\end{tabular}
\caption{The results from this table are obtained by numerical 
integration of Eq. \protect\req{m_ani} for $d=1$ and Monte Carlo
simulations for $d=2$. The existing exact exponent relations
were then applied.}
\end{table}

\vfill\eject
\onecolumn

\begin{figure}
\centerline{\psfig{file=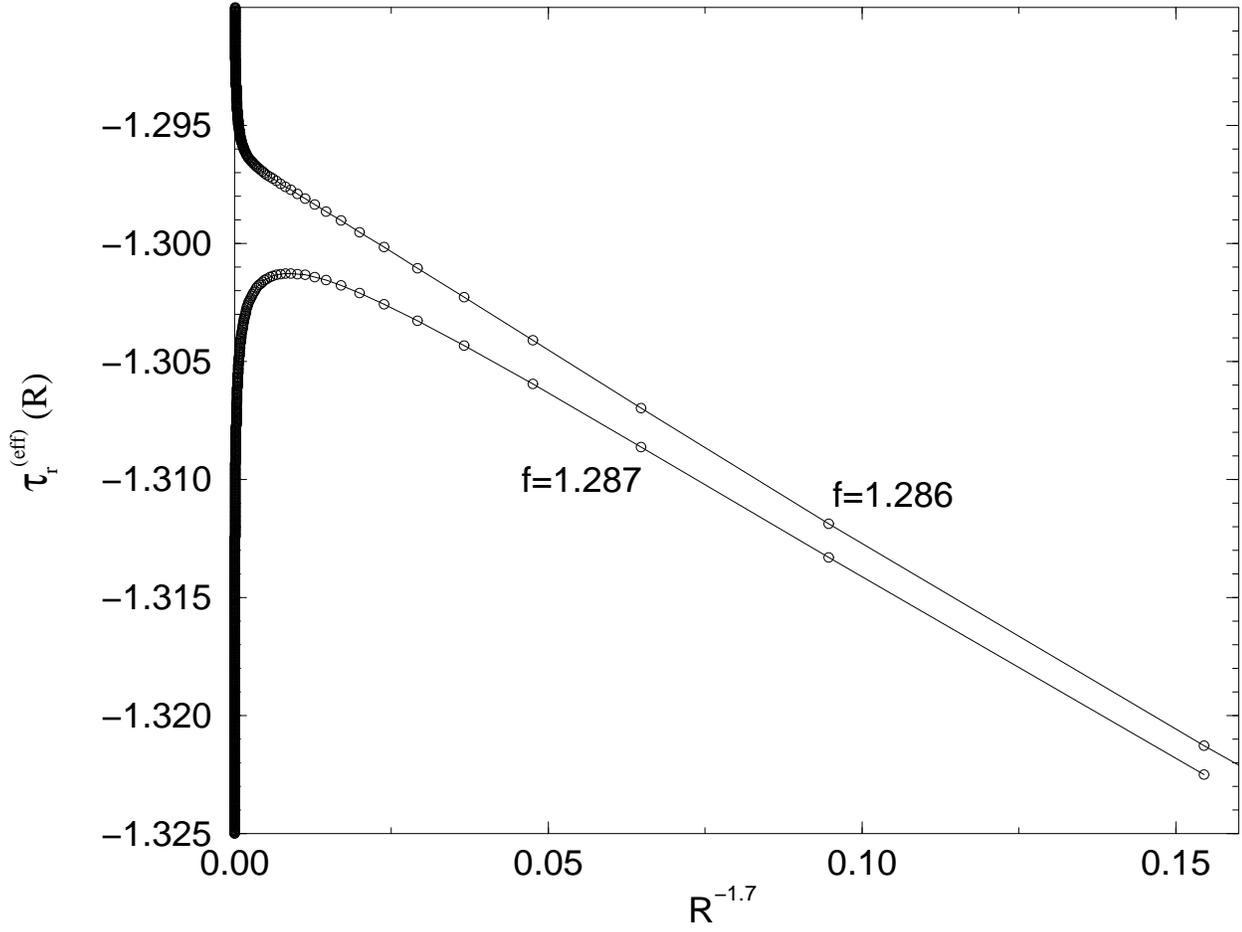,width=\columnwidth,angle=0}}
\caption{The effective power law exponent $\tau_r^{\rm (eff)} (R)$, 
measured as $\tau_r^{\rm (eff)} (R)=
(\log Q(R,f) - \log Q(R-1,f))/(\log R - \log (R-1))$ 
for two different $f$. 
Since our method is free from finite size effects, 
one can be sure that $f_c$ is in $(1.286, 1.287)$ 
and $\tau_r = 1.299(3)$. 
$Q(R,f)$ was obtained by numerical integration of 
\protect\req{m_ani} with $R\le R_{\rm max}=2^{14}=16384$.
A second order Runge Kutta method with $\delta f=10^{-3}$ was used.}

\end{figure}

\begin{figure}
\centerline{\psfig{file=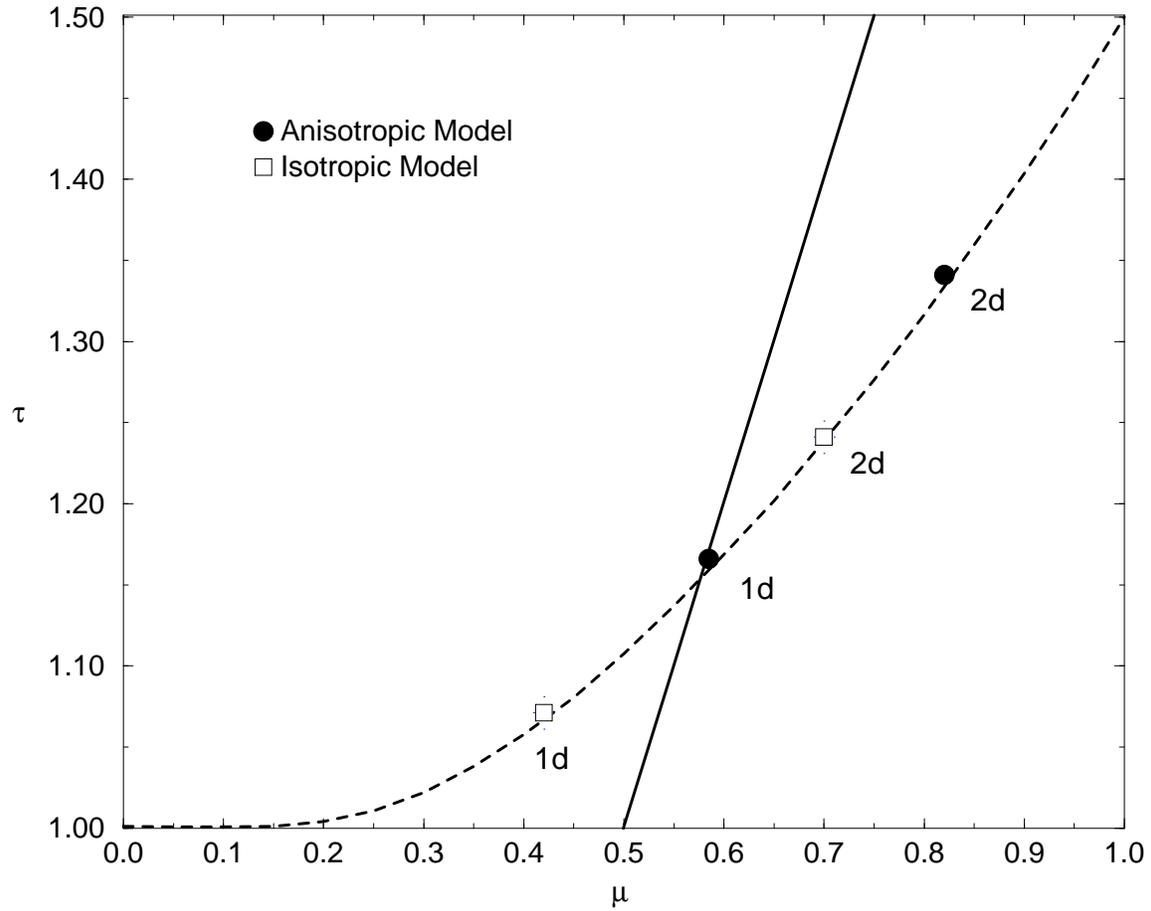,width=\columnwidth,angle=270}}
\caption{The curve $\tau(\mu)$ from Ref. 9
is shown along with the results of 
Monte Carlo simulations of both isotropic ($\circ$) 
and anisotropic ($\bullet$) BS models in
one and two dimensions. The intersection point 
of $\tau(\mu)$ with the line $\tau=2\mu$ 
determines the exponents of the 1D anisotropic 
BS model in agreement with Monte Carlo results.}
\label{fig1}
\end{figure}

\begin{figure}
\centerline{\psfig{file=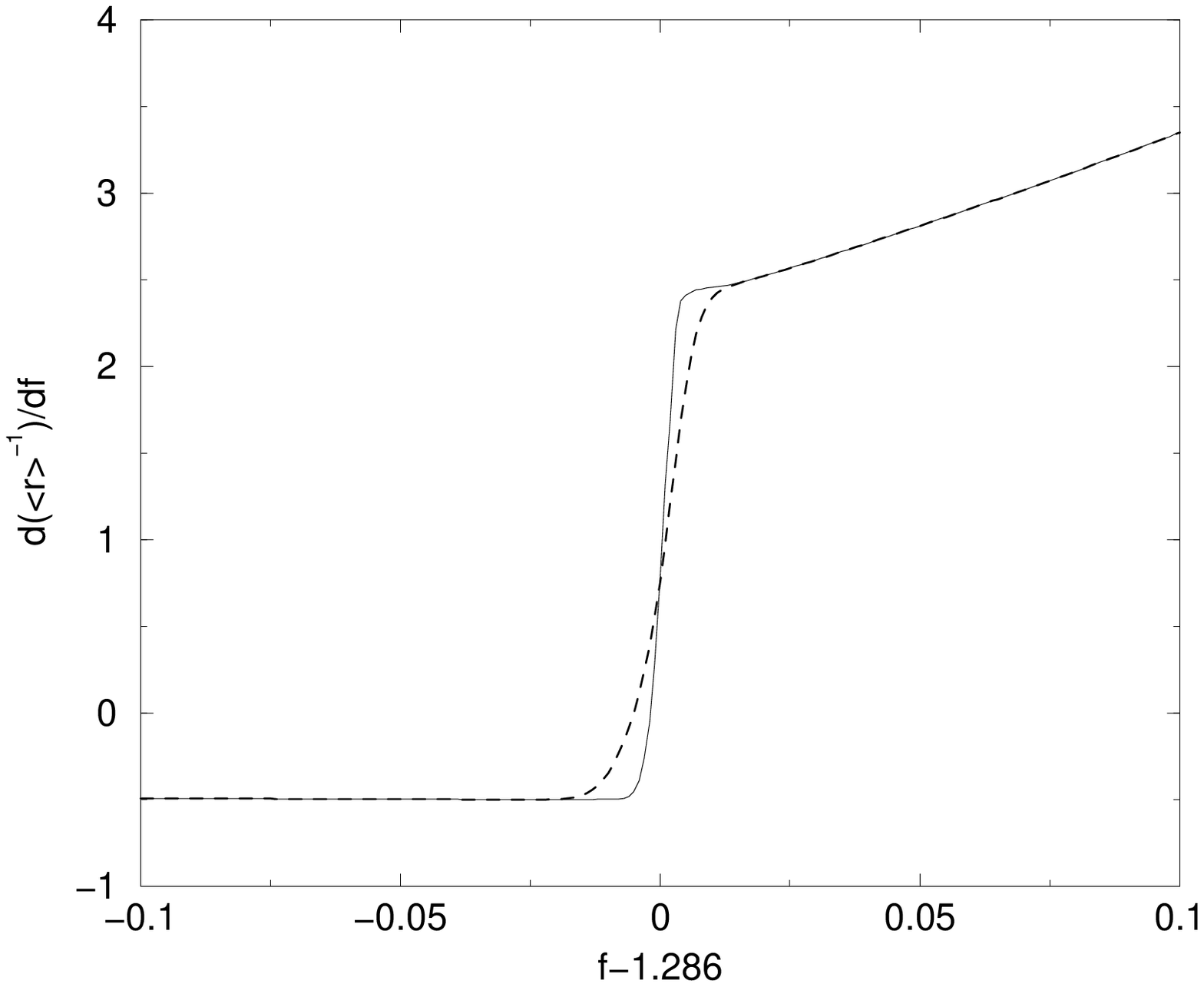,width=\columnwidth,angle=0}}
\caption{The plot of $d(\avg{r}^{-1})/df$ vs $f$ from the numerical
integration of Eq.(3) for $r\le R=2^{12}=1024$ (dashed line) and 
$r\le R=2^{14}=16384$ (solid line). 
A second order Runge-Kutta method with $\delta f=10^{-3}$ was used.}
\label{fig2}
\end{figure}


\begin{references}
\bibitem{notap}
The BS model is ``insensitive''
to the shape of random number distribution
because only the {\it order} among
variables $f_i$ matters in the dynamics. With the 
rescaling map 
$f(f')= -\ln \int_{f'}^{\infty} P(f') df'$,
the BS dynamics for a general $P(f')$ is
turned into that with the exponential 
distribution.
\bibitem{remark} 
Traditionally the Bak-Sneppen model was studied 
with ${\cal P} (f)$ being a flat distribution on $[0,1]$.
In this case one has 
$\avg{r} = \gamma (1-f_c)/(f_c-f)+O((f_c-f)^{-2})$. After 
rescaling of $f_i$ as described in [11] one recovers 
our result for  ${\cal P} (f)=e^{-f}$.
\end{references}
\end{document}